\documentclass[3p,times]{elsarticle}

\usepackage{ecrc}

\usepackage{graphics}

\volume{00}

\firstpage{1}

\journalname{Nuclear Physics A}

\runauth{R.~J.~Fries et al.}

\jid{nupha}

\usepackage{amssymb}
\usepackage[figuresright]{rotating}

\begin{document}

\begin{frontmatter}

\dochead{}
%% Use \dochead if there is an article header, e.g. \dochead{Short communication}

\title{Event-by-Event Jet Quenching and Higher Fourier Moments of Hard Probes}

%% use optional labels to link authors explicitly to addresses:
%% \author[label1,label2]{<author name>}
%% \address[label1]{<address>}
%% \address[label2]{<address>}

\author{Rainer~J.~Fries}
\address{Cyclotron Institute and Department of Physics and Astronomy,
Texas A\&M University, College Station TX 77845, USA}
\address{RIKEN/BNL Research Center, Brookhaven National Laboratory, Upton NY
  11973, USA}

\author{Ricardo Rodriguez}
\address{Department of Mathematics and Physics, Ave Maria University, Ave
  Maria FL 34142, USA}
\address{Cyclotron Institute,
Texas A\&M University, College Station TX 77845, USA}

\begin{abstract}
We investigate the effect of event-by-event fluctuations of the fireball
created in high energy nuclear collisions on hard probe observables.
We show that spatial inhomogeneities lead to changes in the nuclear 
suppression factor of high momentum hadrons which can be absorbed
in the quenching strength $\hat q$. This can increase the theoretical
uncertainty on extracted values of $\hat q$ by up to 50\%. We also
investigate effects on azimuthal asymmetries $v_2$ and dihadron correlation
functions. The latter show a promising residual signal of event-by-event
quenching that might allow us to estimate the size of spatial 
inhomogeneities in the fireball from experimental data.
\end{abstract}

\begin{keyword}
Quark Gluon Plasma \sep Heavy Ion Collisions \sep Energy Loss 

%% MSC codes here, in the form: \MSC code \sep code
%% or \MSC[2008] code \sep code (2000 is the default)

\end{keyword}

\end{frontmatter}

%%
%% Start line numbering here if you want
%%
% \linenumbers

High momentum quarks and gluons are a convenient probe of the quark gluon plasma
bubble formed in high energy nuclear collisions at the Relativistic Heavy Ion
Collider (RHIC) and the Large Hadron Collider (LHC). Those hard partons
are produced in rare hard processes in nuclear collisions. The
process how such quarks and gluons convert into jets of hadrons in the vacuum,
e.g.\ if the they are produced in $p+p$ collisions, is fairly well
documented. On the other hand we are still working to fully understanding how 
the same process plays out in a medium whith which those partons
interact. Over the years several ways to compute the energy loss of energetic
partons in quark gluon plasma have emerged 
\cite{Baier:1996sk,Wiedemann:2000tf,Gyulassy:2000er,Arnold:2002ja}, 
using different approximations to master the full complexity of the problem
(see \cite{Majumder:2010qh,Fries:2010ht} for recent reviews).

Calculations based on these models fare well with data from RHIC on nuclear 
modification factors $R_{AA}$ for leading hadrons if the overall quenching
strength can be adjusted as a parameter. This parameter is often chosen to be
$\hat q$, the average momentum square transferred to the high momentum parton
per mean free path. However, there are large discrepancies between the extracted
values of $\hat q$ even if other details of the calculation, like the
simulation of the space-time evolution of the quark gluon bubble are kept identical
\cite{Bass:2008rv}. Moreover, there is reason to believe that phenomenological
details in the calculation, e.g.\ from fireball evolution or early-time
behavior have a significant effect on extracted values of $\hat q$.

We will argue here that inhomogeneities and fluctuations in the transverse
spatial distribution of both the medium (the quark gluon plasma) and the hard
processes have an effect on $\hat q$ as well. On the other hand, we will show
that precision measurements of hard probes can provide valuable experimental
constraints on such inhomogeneities \cite{Rodriguez:2010di}.

Let us consider a fast parton emerging from a point $\vec r$ in the transverse
plane with zero longitudinal momentum, traveling at a given azimuthal angle
$\psi$ in the transverse plane. For a wide class of jet quenching models the 
average energy loss for such a parton traveling through a medium with local density
$\rho$ is determined by the integral
\begin{equation}  
  I_\beta (\vec r, \psi) = \int d\tau \tau^\beta \rho(\vec r + \tau \vec e_\psi)
\end{equation}
along its trajectory where $\tau$ is the time parameter. The parameter $\beta$ 
determines the length dependence and would be equal to 1 in order to describe
the consequences of the finite formation time known as the LPM effect 
\cite{Majumder:2010qh,Fries:2010ht}.
The impact of energy loss on the parton spectrum as a function of momentum
$p_T$ and angle $\psi$ is governed by the integral over all emission points
$\vec r$ weighted with $n(\vec r)$, the probability density of jets
emerging at point $\vec r$,
\begin{equation}
  \int d^2 r \int d\tau  \tau^\beta n(\vec r) \rho(\vec r + \tau \vec e_\psi)
\end{equation}
For the case of small energy loss $\Delta p_T \ll p_T$ the correction to the
spectrum is directly 
proportional to this integral. For all other cases the relation might be more
complicated, but it is still a monotonous function of
the density product $n(\vec r) \rho(\vec r + \tau \vec e_\psi)$.

\begin{figure}[tb]
\centering%
\includegraphics[width=65mm]{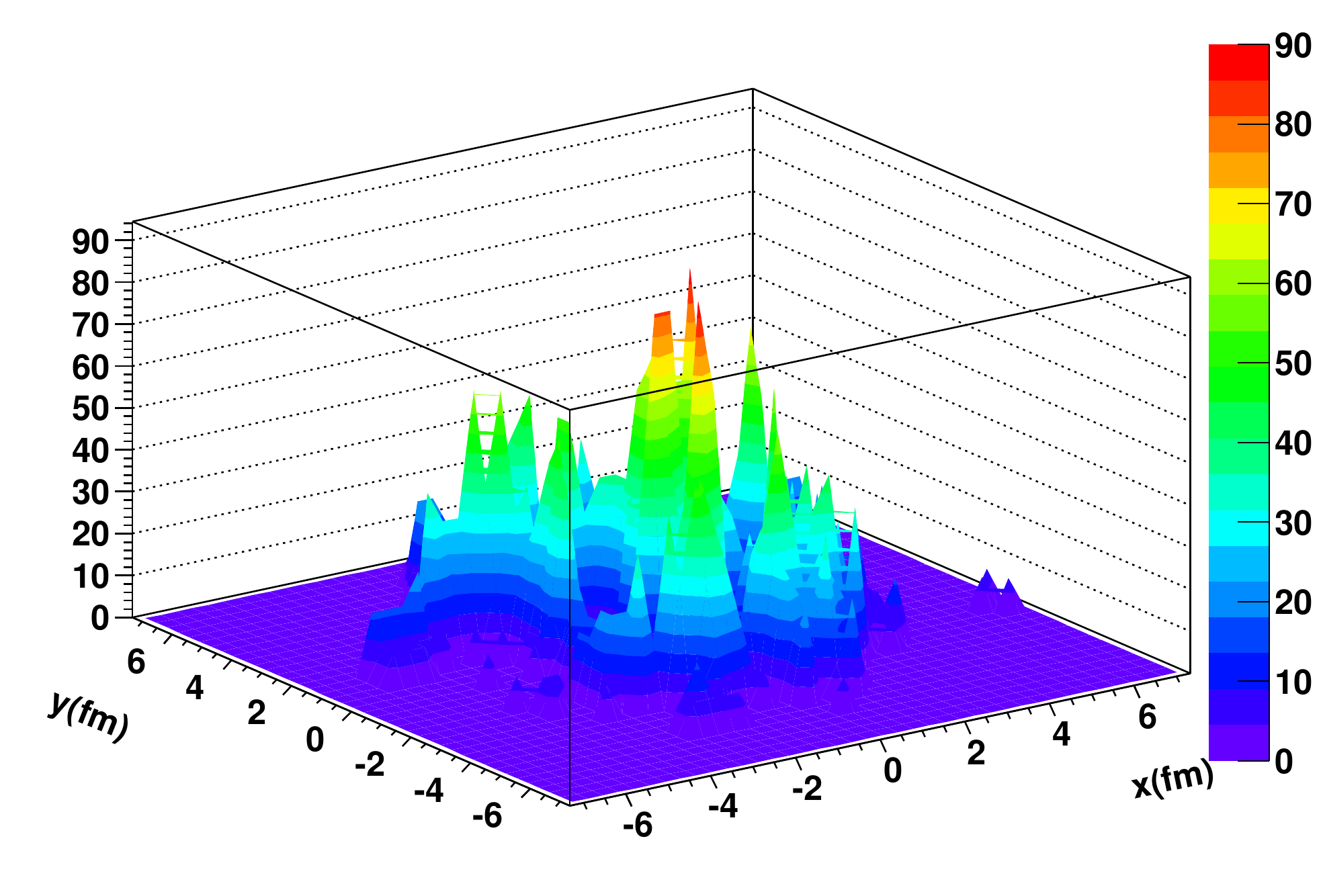} \,
\includegraphics[width=65mm]{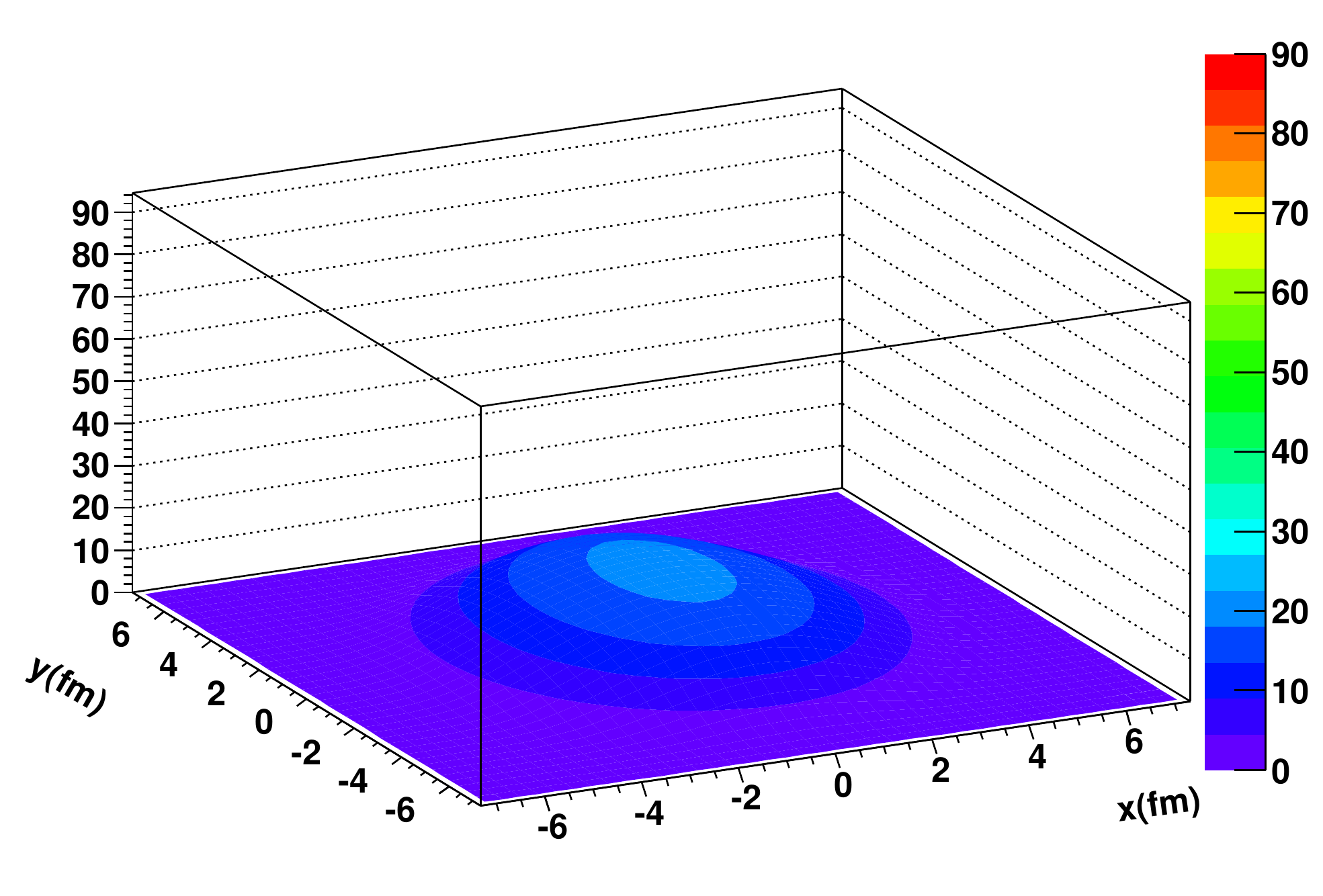}
\caption{Density of nucleon-nucleon collisions in the transverse plane in
  Au+Au collisions  with average impact parameter $b=3.2$ fm generated with
  GLISSANDO.  Left panel: A random single event. 
  Right panel: Average over 500,000 events.}
\label{fig:1}
\end{figure}

Now let us assume that the densities $n(\vec r)$ and $\rho(\vec r)$ fluctuate 
event-by-event around expectation values $\bar n(\vec r)$ and $\bar \rho(\vec r)$:
$n = \bar n + \delta n$, $\rho = \bar \rho + \delta \rho$. The effect on
quenching from the density product, averaged over many events, can then be
broken down into two terms
\begin{equation}
  \langle n(\vec r) \rho(\vec r + \tau \vec e_\psi)  \rangle = \bar n(\vec r) \bar 
  \rho(\vec r + \tau \vec e_\psi) + R(\vec r, \vec r+\tau \vec e_\psi)
\end{equation}
where the first term corresponds to quenching in a smooth, averaged event,
while the deviations from it are described by the correlation function
\begin{equation}
  R(\vec r_1, \vec r_2) = \langle \delta n(\vec r_1) \delta \rho(\vec r_2)
  \rangle
\end{equation}
In other words deviations of quenching when correctly calculated
event-by-event and compared to averaged fireballs are sensitive to the
correlation between the density of jet emissions and the density of the medium
along the path of the jet.

\begin{figure}[tb]
\centering%
\includegraphics[width=70mm]{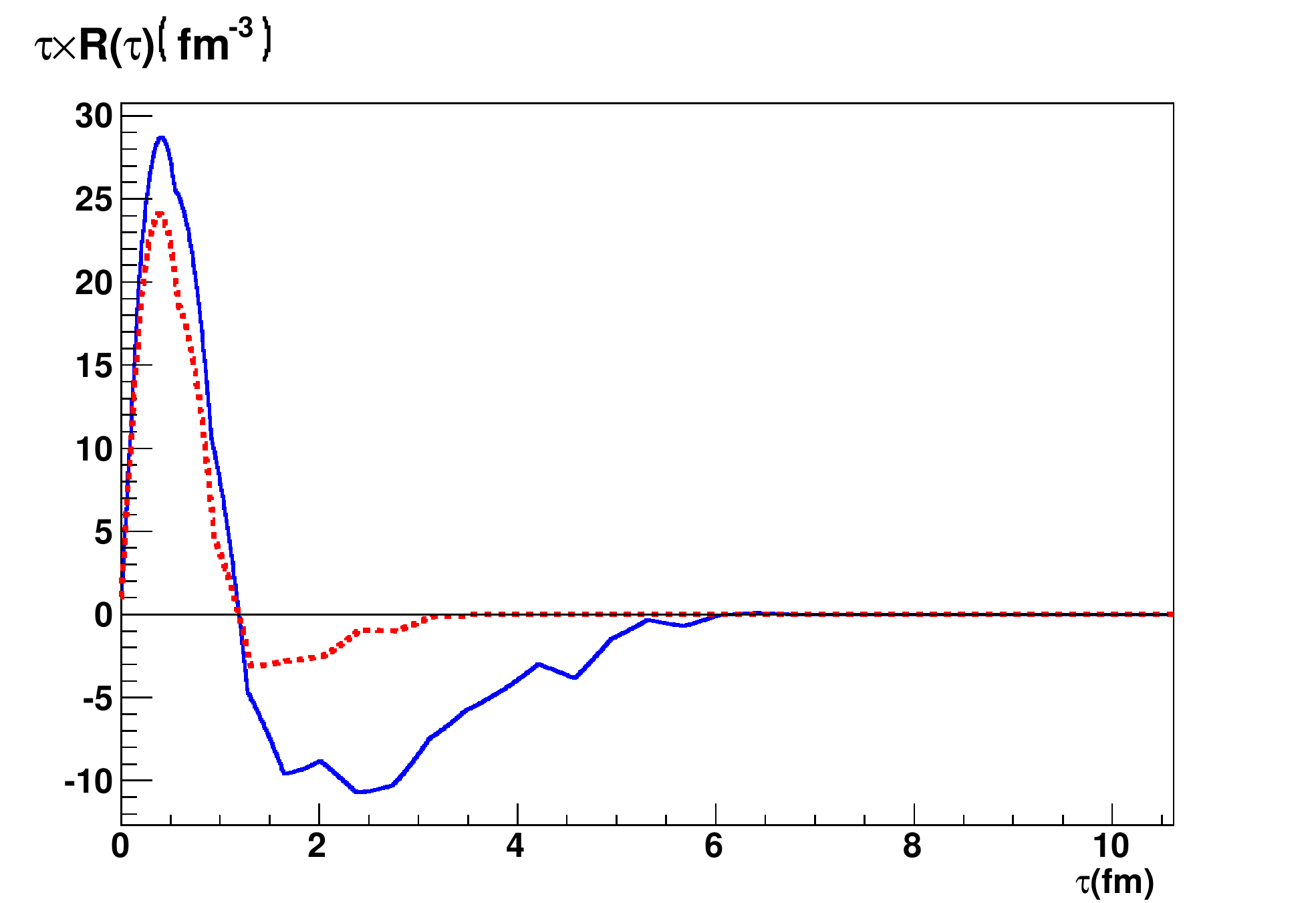} \,
\caption{The correlation function $\tau R(\tau)$ from a position $\vec r= 4$
fm $\vec e_y$ ($\vec r=0$ corresponds to the center of the fireball)
in the directions $\psi=0$ (solid line) and $\psi=\pi/2$ (dashed line) in
central Au+Au collisions calculated with GLISSANDO.}
\label{fig:2}
\end{figure}

If we were able to measure these deviations we could gain access to the
correlation function $R$, and hence estimate the average spatial
inhomogeneities in the fireball, despite the fact that high transverse
momentum ($P_T$) data at RHIC is averaged over many events. 
E.g.\ $R$ would tell us about the
average size of the typical granularity in events or the size of so-called hot spots. 
Fig.\ \ref{fig:1} shows the distribution $n(\vec r)$ represented by the number
of nucleon-nucleon collisions for a typical Au+Au collision with small impact
parameter ($b\approx 3.2$ fm) at RHIC energies together with the average
distribution $\bar n(\vec r)$. The distributions were calculated with the
Glauber Monte Carlo GLISSANDO \cite{Broniowski:2007nz}. Fig.\ \ref{fig:2}
shows the correlation function $R(\tau) = R(\vec r, \vec r+\tau e_\psi)$
calculated with GLISSANDO events. One can clearly see a region of positive
correlation around the emission point (small $\tau$) and a tail of
anti-correlation extending to the boundary of the fireball.

This behavior of $R$ confirms the intuitive expectations. The integrals over
$\tau$ and $\vec r$ will pick up positive and negative contributions to the
correlation function and it is hard to argue whether the net effect on the
spectra is an enhancement or suppression of quenching compared to the
case of average fireballs. However, interestingly one can argue that the
azimuthal asymmetry $v_2$ should decrease since the anti-correlation
tails extend all the way to the boundary of the fireball
\cite{Rodriguez:2010di}.

We have run numerical simulations using our PPM package
\cite{Rodriguez:2010di}
utilizing two different energy loss models: (i) a simple
LPM-inspired model (sLPM) with $\Delta E = c_{\rm sLPM} I_1(\vec r, \psi)$ and
(ii) the non-deterministic ASW/BDMPS model \cite{Wiedemann:2000tf}. In both
cases we model the densities $n$ and $\rho$ using the binary nucleon collision
densities from GLISSANDO. Note that the parameters $c_{\rm sLPM}$ and
$c_{\rm ASW}$ can be interpreted as the local quenching strength per density,
$\hat q/\rho$.

\begin{figure}[tb]
\centering%
\includegraphics[width=75mm]{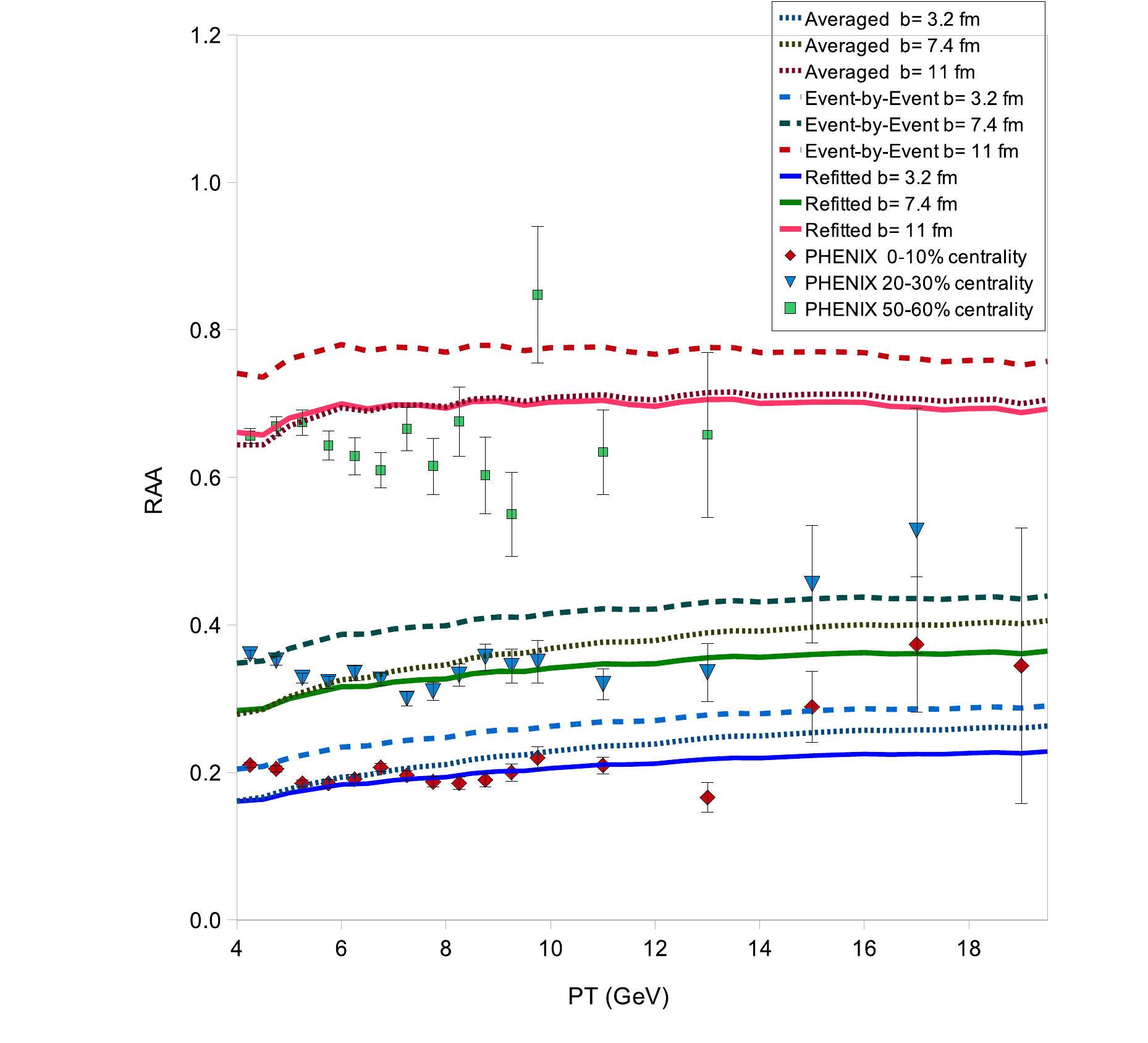} \,
\includegraphics[width=80mm]{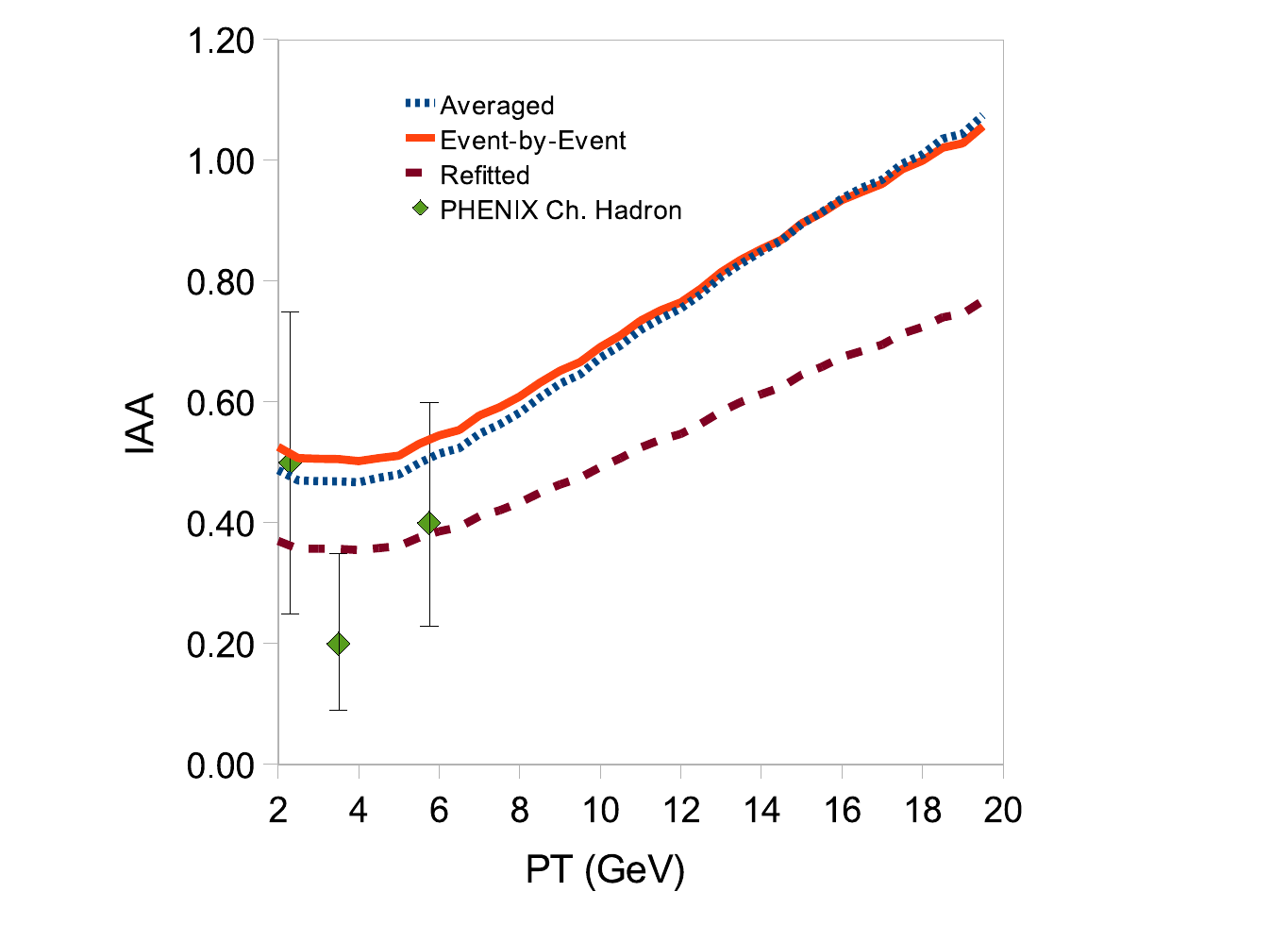}
\caption{Left panel: Nuclear modification factor $R_{AA}$ for neutral pions in
  Au+Au collisions in three
  different centrality bins. Data from PHENIX \cite{Adare:2008qa} is compared
  to calculations using averaged events with $\hat q$ fitted (dotted
  lines), event-by-event calculations with the same $\hat q$ (dashed lines),
  and event-by-event calculations with $\hat q$ refitted
  to the data (solid lines). Right panel:
  Triggered $\pi^0$-$\pi^0$ correlation $I_{AA}$ for central Au+Au
  collisions as a function of associated particle momentum $P_T$. The trigger
  particle momentum lies between 7 and 9 GeV. Data from the PHENIX
  Collaboration \cite{Adare:2010ry} is compared to our calculations for the
  averaged event (dotted line), the event-by-event calculations
  with unchanged $\hat q$ (solid line) and the event-by-event calculation with $\hat q$
  refitted to describe $R_{AA}$. 
}
\label{fig:3}
\end{figure}

Fig.\ \ref{fig:3} shows results for $R_{AA}$ of neutral pions obtained with
the ASW model for three different centralities compared to data from PHENIX  
\cite{Adare:2008qa}. First $c_{\rm ASW}$ is fitted using averaged events
without inhomogeneities (``Averaged''), then the same value of $c_{\rm ASW}$ is
used in an event-by-event calculation (``Event-by-Event''). We clearly see a 
reduction in quenching for all centralities and momenta going from averaged
events to realistic, fluctuating events. However, we can adjust $c_{\rm ASW}$
from 1.6 GeV to 2.8 GeV to refit the $R_{AA}$ data on the same level of
accuracy in the event-by-event case (``Refitted''). Calculations using the
sLPM energy loss model yield similar results \cite{Rodriguez:2010di}.
We conclude that in absence of any knowledge about the spatial inhomogeneities
in the fireball this introduces an additional uncertainty on the extracted
value of $\hat q$ (or any version of this quantity normalized to a density)
of at least 50\%.

The next interesting question is whether after refitting $R_{AA}$  any
observable consequences of spatial inhomogeneities remain. We find that
$v_2$ at high $P_T$ is decreased going from averaged events to an 
event-by-event calculation with the same $\hat q$, as predicted above. With
the quenching strength used to refit $R_{AA}$ the results for $v_2$ are still slightly
below the results in averaged events, but the deviations (typically less than
20\%) might be too subtle to be useful. On the other hand, triggered two
particle correlations might be sensitive enough to put constraints on $R$.
When going from averaged to event-by-event calculations using the same $\hat
q$ the decreased suppression almost cancels between hadron pairs and trigger
particles, leading to very similar curves. After readjusting $R_{AA}$ we find
a rather large suppression of $I_{AA}$ as shown in Fig.\ \ref{fig:3}.

It is also interesting to ponder the effects that inhomogeneities can have on
azimuthal asymmetry coefficients other than $v_2$.  We have analyzed 
a set of engineered events with given spatial anisotropies and have found that
the Fourier coefficients $v_n$ scale linearly with the (generalized) spatial
eccentricities $\epsilon_n$ in both the ASW and the sLPM energy loss models
for $n=2$, 3, 4. Using samples of GLISSANDO events we have indeed found 
Fourier coefficients up to $n=6$. However except for $v_2$ and $v_4$ the magnitude
of these coefficients is generally below 1\%. Further details will be reported
in a forthcoming publication \cite{Fries:2011ta}.

Let us summarize. Inhomogeneities in the space-time structure of a quark gluon
fireball have a potentially large effect on hard probes. One can
define a correlation function $R$ which encodes valuable information on the 
granularity and average magnitude of these fluctuations and which can
potentially be extracted even in event-averaged measurements in heavy ion
collisions. We have studied the function $R$ within the GLISSANDO Glauber
model and have investigated the effect of event-by-event quenching for two
different energy loss models. We find that $R_{AA}$ increases and $v_2$
decreases event-by-event compared to averaged events if the same quenching 
strength $\hat q$ per density is used. However, we can not use this effect to 
constrain $R$ since a simple redefinition of $\hat q$ can fit RHIC data quite
well. However, in turn this means that there is a roughly 50\%
additional uncertainty on the extracted values of $hat q$ which comes from our lack of 
knowledge about $R$. On the other hand we 
observe that two particle correlations like $I_{AA}$ carry residual signatures
of the correlation function $R$. If all other uncertainties were under control
they could be used to experimentally constrain $R$ and gain tomographic 
insight into the spatial structure of the fireball.
We have also found a linear scaling of $v_n$ with $\epsilon_n$ at large
momentum and we find higher order Fourier coefficients in event-by-event
jet quenching calculations which might be accessible in future measurements.

Acknowledgments: This work was supported by CAREER Award PHY-0847538
from the U.S.\ National Science Foundation, RIKEN/BNL and DOE grant 
DE-AC02-98CH10886.

%% References with BibTeX database:

\bibliographystyle{elsarticle-num}
\bibliography{Fries_HP10}

\end{document}